\documentclass{aipproc}
\layoutstyle{6x9}
\usepackage{graphicx}
\usepackage{amssymb}

\begin{document}

\newcommand {\e} {\varepsilon}
\newcommand {\ph} {\varphi}

\title{Synchronization of Limit Cycle\\Oscillators by Telegraph Noise}

\classification{05.40.-a; 02.50.Ey; 05.45.Xt}
\keywords{Noise; Synchronization; Lyapunov Exponent}
\author{Denis S. Goldobin}
{address={Department of Physics, University of Potsdam,
          Postfach 601553, D-14415 Potsdam, Germany},
 address={Department of Theoretical Physics, Perm State University,
             15 Bukireva str., 614990, Russia}}

\begin{abstract}
We study the influence of telegraph noise on synchrony of limit
cycle oscillators. Adopting the phase description for these
oscillators, we derive the explicit expression for the Lyapunov
exponent. We show that either for weak noise or frequent switching
the Lyapunov exponent is negative, and the phase model gives
adequate analytical results. In some systems moderate noise can
desynchronize oscillations, and we demonstrate this for the Van
der Pol--Duffing system.
\end{abstract}

\maketitle

\section{Introduction}

In autonomous systems exhibiting a stable periodic behavior (in
other words, limit cycle oscillators), deviations along the
trajectory asymptotically in time neither decay nor grow, i.e.
they are neutral. This neutrality is due to the time homogeneity,
and may disappear as soon as this homogeneity is broken by means
of a time-dependent external forcing. The phenomenon of
synchronization of oscillators by periodic signal is well known
and quite understood, here the oscillations follow the forcing
(e.g., they attain the same frequency). When the role of this
time-dependent forcing is played by a stochastic noise, the
situation becomes less evident.

The first effect of noise on periodic oscillations is the phase
diffusion: the oscillations are no more periodic but posses finite
correlations~\cite{Stratonovich-63, Malakhov-68}. However, a
nonlinear system may somehow follow the noisy force. Although it
is not so evident how the synchronization between the system
response and noise can be detected for one system, this
synchronization can be easily detected by looking on whether the
responses of a few identical systems driven by the common noise
signal are identical or not. With such an approach, synchrony
(asynchrony) of driven systems was early treated in the
works~\cite{Pikovsky-1984a, Pikovsky-1984b,
Pikovsky-Rosenblum-Kurths-2001, Yu-Ott-Chen-90,
Baroni-Livi-Torcini-2001, Khoury-Lieberman-Lichtenberg-96}.
 The mathematical criterion for synchronization is the negative
leading Lyapunov exponent (LE; it measures the average exponential
growth rate of infinitesimally small deviations from the
trajectory) in the driven system.

In different fields the effect of synchronization of oscillators
by common noise is known under different names. In neurophysiology
the property of a single neuron to provide identical outputs for
repeated noisy input is treated as
"reliability"~\cite{Mainen-Sejnowski-95}. In the experiments with
noise-driven Nd:YAG lasers~\cite{Uchida-Mcallister-Roy-2004} this
synchronization was called "consistency". When driving signal is
related to not stochastic but deterministic chaos, one considers
generalized synchronization~\cite{Abarbanel-Rulkov-Sushchik-96}.
In the last case the driven system is often chosen to be chaotic.
In fact, in the above mentioned examples there is no limit cycle
oscillators at the noiseless limit: for experiments described
in~\cite{Mainen-Sejnowski-95, Uchida-Mcallister-Roy-2004} the
noiseless system is stable, i.e. LE is negative, and for
generalized synchronization in chaotic systems, LE is positive.
Evidently, in this cases, LE preserves its sign at sufficiently
weak noise.

A more intriguing situation takes place when LE in noiseless
system is zero (limit cycle oscillators). Analytical and numerical
treatments for different types of noise show weak noise to play an
ordering role: LE shifts to negative values, and oscillators
become synchronized~\cite{Pikovsky-1984a, Pikovsky-1984b,
Pikovsky-Rosenblum-Kurths-2001, Teramae-Tanaka-2004,
Goldobin-Pikovsky-2005a}. In the
work~\cite{Goldobin-Pikovsky-2005b} the nonideal situations are
considered: slightly nonidentical oscillators driven by an
identical noise signal, and identical oscillators driven by
slightly nonidentical noise signals; and additionally, positive LE
was reported for large noise in some smooth systems similarly to
how it was in the works~\cite{Pikovsky-1984a, Pikovsky-1984b}.

Note that in~\cite{Pikovsky-1984a, Pikovsky-1984b} LE was
calculated for oscillators driven by a random sequence of pulses,
in~\cite{Pikovsky-Rosenblum-Kurths-2001, Teramae-Tanaka-2004,
Goldobin-Pikovsky-2005a, Goldobin-Pikovsky-2005b} the white
Gaussian noise was considered. A noise of other nature is the
telegraph one. By a normalized telegraph noise we mean the signal
having values $\pm1$ and switching instantaneously between these
values time to time. The distribution of time intervals between
consequent switchings is exponential with the average value
$\tau$. The case of telegraph noise may be interesting not only
because it completely differs from the previous two, but also
because it allows to "touch" the question of relations between
periodic and stochastic driving, e.g. to compare results for
telegraph noise with the average switching time $\tau$ and the
stepwise periodic signal of the same amplitude and the period
$2\tau$. This is why we consider synchronization by telegraph
noise.

\section{Phase Model}

A limit cycle oscillator with a small external force is known to
be able to be well described within the phase
approximation~\cite{Kuramoto-1984-2003}, where only dynamics of
the system on the limit cycle of the noiseless system is
considered\footnote{Noteworthy, the phase approximation is valid
not only for a small external force, but also for a moderate one
if only the leading Lyapunov exponent of the limit cycle is
negative and large enough.}. The system states on this limit cycle
can be parameterized by a single parameter, phase $\ph$. With a
stochastic force the equation for the phase reads
\begin{equation}
\dot{\ph}=\omega+\e f(\ph)\xi(t)\;,\label{e1}
\end{equation}
where $2\pi/\omega$ is the period of the limit cycle in the
noiseless system, $\e$ is the amplitude of noise, $f(\ph)$ is the
normalized sensitivity of the system to noise
[~$(2\pi)^{-1}\int_0^{2\pi}f^2(\ph)d\ph=1$], and $\xi$ is a
normalized telegraph noise.

\subsection{Master equation}
Studying statistical properties of the system under consideration,
one can introduce two probability density functions
$W_{\pm}(\ph,\,t)$ defining the probability to locate the system
in vicinity of $\ph$ with $\xi=\pm1$, correspondingly, at the
moment $t$. Then the Master equations of the system read
\begin{eqnarray}
\frac{\partial W_+(\ph,\,t)}{\partial t}
 +\frac{\partial}{\partial\ph}\left[(\omega+\e f(\ph))W_+(\ph,\,t)\right]&=&
\frac{1}{\tau}W_-(\ph,\,t)-\frac{1}{\tau}W_+(\ph,\,t),\label{e2}\\
\frac{\partial W_-(\ph,\,t)}{\partial t}
 +\frac{\partial}{\partial\ph}\left[(\omega-\e f(\ph))W_-(\ph,\,t)\right]&=&
\frac{1}{\tau}W_+(\ph,\,t)-\frac{1}{\tau}W_-(\ph,\,t).\label{e3}
\end{eqnarray}
In the terms of $W\equiv W_++W_-$ and $V\equiv W_+-W_-$ the last
system takes the form of
\begin{equation}
\dot{W}=-\omega W_{\ph}-\e\left(f\,V\right)_{\ph},\qquad
\dot{V}=-\omega V_{\ph}-\e\left(f\,W\right)_{\ph}
-\frac{2}{\tau}V.\label{e4}
\end{equation}

For steady distributions the probability flux $S$ is constant:
\[S=\omega W(\ph)+\e f(\ph)\,V(\ph);\]
and system~(\ref{e4}) with periodic boundary conditions has the
solution
\begin{equation}
V(\ph)=-\frac{\e\omega\,C}{\omega^2-\e^2f^2(\ph)}
\int\limits_{\ph}^{\ph+2\pi}d\psi\,f'(\psi)
\exp\left(\frac{2}{\tau}\int\limits_{\ph}^{\psi}
\frac{d\,\theta}{\omega^2-\e^2f^2(\theta)}\right),
\label{e5}
\end{equation}
where $C$ is defined by the normalization condition:
\begin{eqnarray}
C^{-1}&=&2\pi\left(\exp\left(\frac{2}{\tau}\int\limits_0^{2\pi}
\frac{d\,\theta}{\omega^2-\e^2f^2(\theta)}\right)-1\right)\nonumber\\
&&+\e^2\int\limits_0^{2\pi}d\ph\int\limits_{\ph}^{\ph+2\pi}d\psi
\frac{f(\ph)\,f'(\psi)}{\omega^2-\e^2f^2(\ph)}
\exp\left(\frac{2}{\tau}\int\limits_{\ph}^{\psi}\frac{d\,\theta}
{\omega^2-\e^2f^2(\theta)}\right).\label{e6}
\end{eqnarray}
The probability flux reads
\[S=\omega\left(\exp\left(\frac{2}{\tau}\int\limits_0^{2\pi}
\frac{d\,\theta}{\omega^2-\e^2f^2(\theta)}\right)-1\right)C.\]

\subsection{Lyapunov exponent}
Studying stability of solutions of the stochastic
equation~(\ref{e1}), one has to consider behavior of a small
perturbation $\alpha$:
\[\dot{\alpha}=\e f'(\ph)\alpha\,\xi(t).\]
The Lyapunov exponent (LE) measuring the average exponential
growth rate of $\alpha$ can be obtained by averaging the
corresponding velocity
\begin{eqnarray}
\lambda&=&\langle\frac{d}{dt}\ln{\alpha}\rangle
 =\langle\e f'(\ph)\xi(t)\rangle
 =\e\int\limits_0^{2\pi}f'(\ph)V(\ph)d\ph\nonumber\\
&=&-\e^2\omega\,C\int\limits_0^{2\pi}d\ph
\int\limits_{\ph}^{\ph+2\pi}d\psi\,
\frac{f'(\ph)\,f'(\psi)}{\omega^2-\e^2f^2(\ph)}
\exp\left(\frac{2}{\tau}\int\limits_{\ph}^{\psi}
\frac{d\,\theta}{\omega^2-\e^2f^2(\theta)}\right).\label{e7}
\end{eqnarray}

Let us remind that LE determines the asymptotic behavior of small
perturbations, and describes whether close states diverge or
converge with time. This process is not necessarily monotonous,
i.e., close trajectories can diverge at some time intervals while
demonstrating asymptotic convergence, and vice versa.

When $\tau\ll 1$ or $\e\ll\omega$, the eq.~(\ref{e7}) can be
simplified:
\begin{equation}
\lambda_{\mathrm{app}} =-\frac{\e^2}{2\pi\omega}
\left(\exp\left(\frac{4\pi}{\tau\omega^2}\right)-1\right)^{-1}
\int\limits_0^{2\pi}d\ph
\int\limits_0^{2\pi}d\psi\,f'(\ph)f'(\psi+\ph)\,
\exp{\frac{2\psi}{\tau\omega^2}}.\label{e8}
\end{equation}
The last expression is strictly negative. Indeed, in the Fourier
space it reads
\[\lambda_{\mathrm{app}}
=-\omega\tau\e^2\sum\limits_{k=1}^{\infty}
 |C_k|^2\frac{k^2}{1+(k\tau\omega^2/2)^2},\]
where $C_k=(2\pi)^{-1}\int\limits_0^{2\pi}f(\ph)\,e^{-ik\ph}d\ph$.

\section{Comparison to Numerical Simulation}

We found that either for weak noise or frequent switching LE is
negative regardless to the properties of the smooth function
$f(\ph)$ (as for weak white Gaussian noise in similar
systems~\cite{Teramae-Tanaka-2004, Goldobin-Pikovsky-2005a}). In
the works~\cite{Goldobin-Pikovsky-2005b}, moderate white Gaussian
noise was shown to be able to lead to instability even in smooth
systems. In the light of above facts, it is interesting (i) {\em
what is the region of validity of our analytical theory}, (ii)
{\em whether there is some footprints of the synchronization by
periodic forcing in the stochastic synchronization}, and (iii)
{\em whether telegraph noise can desynchronize oscillators}.

For the two first purpose we performed simulation of a modified
Van der Pol oscillator:
\begin{equation}
\ddot{x}-\mu(1-x^2-\dot{x}^2)\dot{x}+x=\e\sqrt{2}\xi(t),\label{e9}
\end{equation}
where $\xi(t)$ is either a telegraph noise with the average
switching time $\tau$ or a periodic stepwise signal with the
period $2\tau$, i.e. the constant switching time $\tau$. The
forcing-free modified Van der Pol oscillator has the round stable
limit cycle of the unit radius for all $\mu>0$. Nevertheless, the
phase equation~(\ref{e1}) with $\omega=1$ and the simple function
$f(\ph)=\sqrt{2}\cos{\ph}$ may be correctly adopted only if the
phase speed is near-constant all over the limit cycle, which is
possible at small $\mu$ only.

\begin{figure}
  \centerline{\includegraphics[width=0.62\textwidth]
{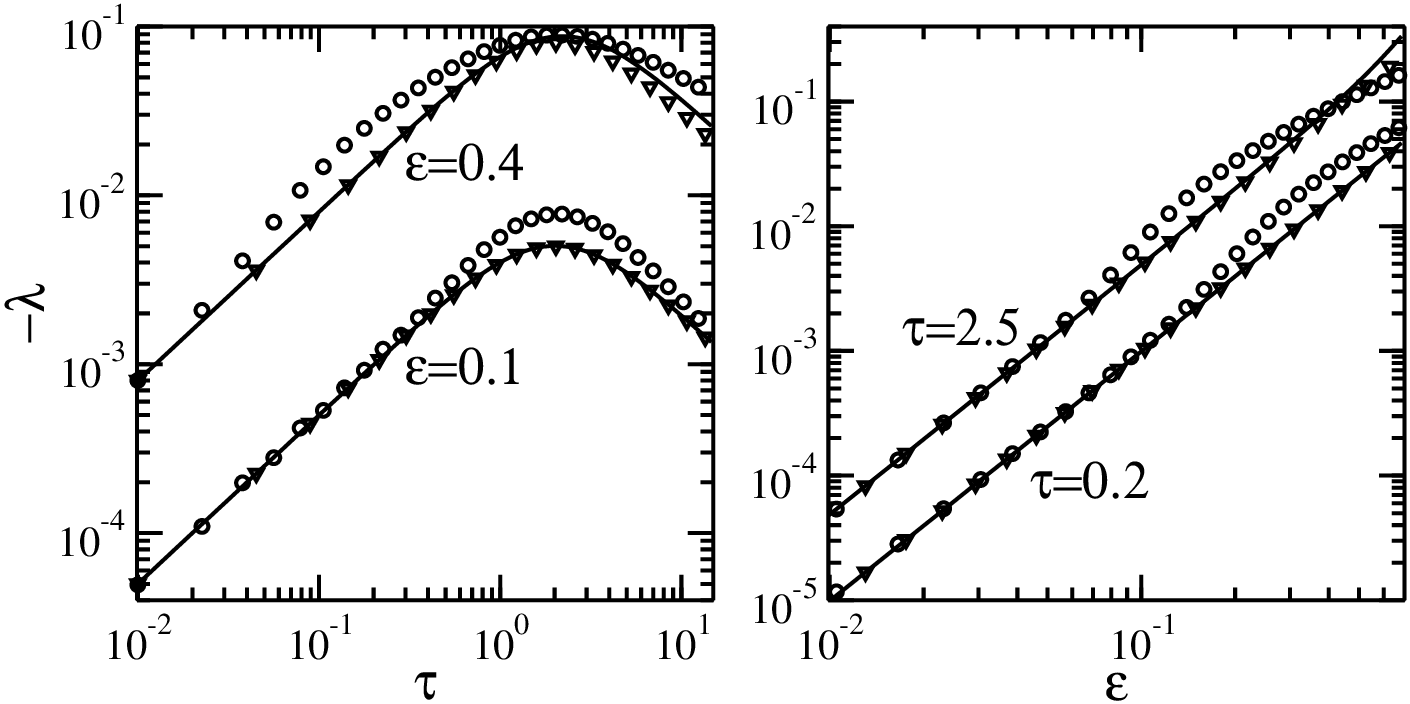}
  \quad\includegraphics[width=0.322\textwidth]
{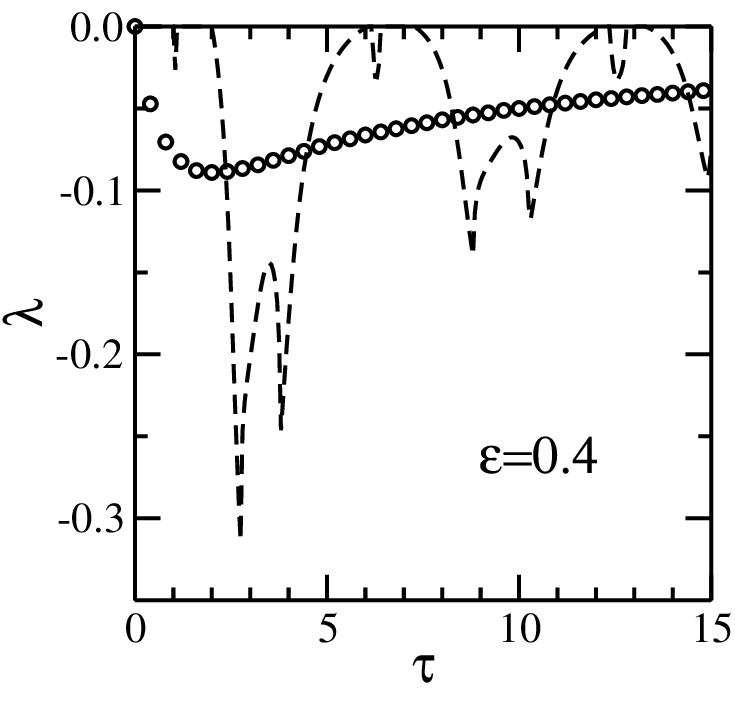}\quad}
  \caption{Samples of dependencies $\lambda(\e,\,\tau)$ for the modified
Van der Pol oscillator~(\ref{e9}) at $\mu=0.1$. The solid lines
present the analytical results of phase description, the triangles
plot results of the approximation~(\ref{e8}), the circles
correspond to numerical simulation of the noisy modified Van der
Pol oscillator, and the dashed line corresponds to numerical
simulation of the periodically driven one.}
  \label{fig1}
\end{figure}

In Fig.~\ref{fig1} one can see that our analytical theory is
fortunately in good agreement with the results of analytical
simulation not only for weak noise; and the dependence
$\lambda(\e,\,\tau)$ for the stochastic driving has no footprints
of the one for the periodic driving.

\begin{figure}
  \centerline{\includegraphics[width=0.95\textwidth]
{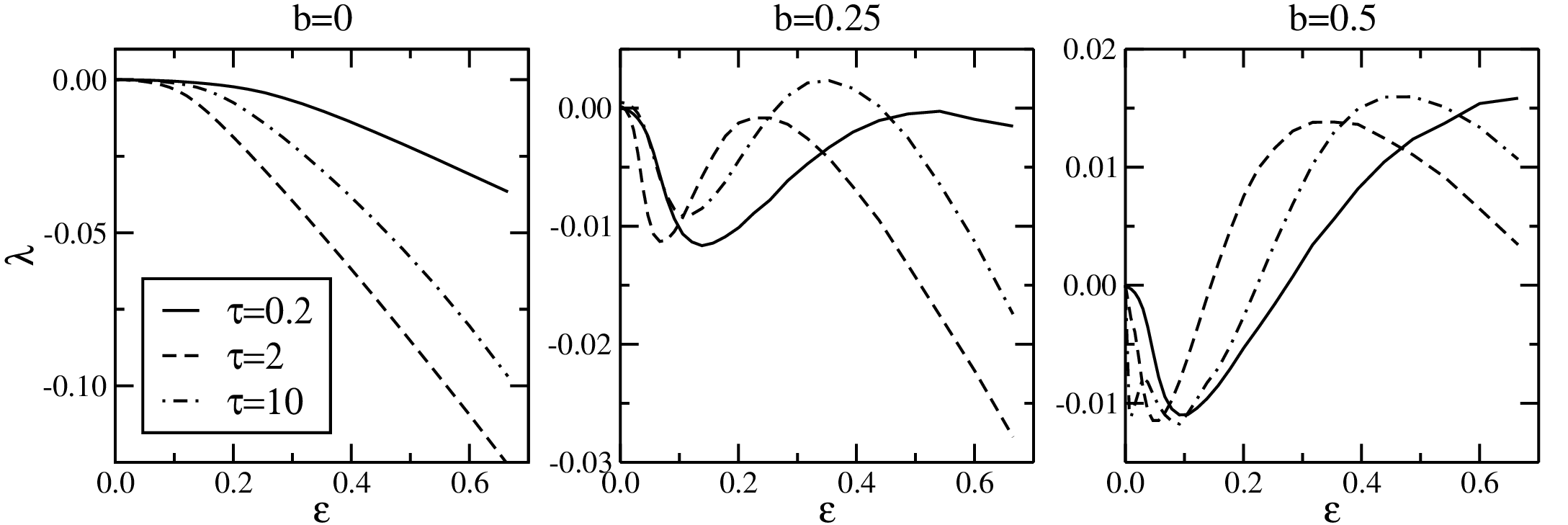}\quad}
  \caption{Samples of dependencies $\lambda(\e,\,\tau)$ for the
Van der Pol--Duffing oscillator~(\ref{e10}) at $\mu=0.1$. The
values of the parameters $b$ and $\tau$ are indicated above the
plots.}
  \label{fig2}
\end{figure}

While the dynamical system~(\ref{e9}) does not exhibit positive
LEs at any noise intensity and any $\mu$, they can be observed for
a Van der Pol--Duffing model\footnote{A similar situation occurs
for white Gaussian noise~\cite{Goldobin-Pikovsky-2005b}.}:
\begin{equation}
\ddot{x}-\mu(1-2x^2)\dot{x}+x+2bx^3=\e\sqrt{2}\xi(t),\label{e10}
\end{equation}
where "Duffing parameter" $b$ describes nonisochronicity of
oscillations. In Fig.~\ref{fig2} one can see that at large enough
$b$ positive LE appears in a certain range of parameters.

\section{Conclusions}

Having considered the phenomenon of synchronization of limit cycle
oscillators by common telegraph noise, we can summarize:

{\em
\noindent --- Either for weak noise or frequent switching the
Lyapunov exponent is negative;

\noindent --- For some systems, the phase model gives quite
adequate results even for moderate noise levels and values of the
average switching time;

\noindent --- The dependence $\lambda(\e,\,\tau)$ for stochastic
driving does not look to have any footprints of the one for
periodic driving;

\noindent --- In some systems, moderate telegraph
noise can desynchronize oscillations.}

Here we do not present results for the nonideal situations (like
in~\cite{Goldobin-Pikovsky-2005b}): slightly nonidentical
oscillators driven by an identical noise signal, and identical
oscillators driven by slightly nonidentical noise signals. The
reason is that for weak noise these results appear to be the same
as in~\cite{Goldobin-Pikovsky-2005b} but with
$\lambda_{\mathrm{app}}$ given by Eq.~(\ref{e8}) instead of
$\lambda$.

\begin{theacknowledgments}
The work has been supported by DFG (SFB 555).
\end{theacknowledgments}

\bibliographystyle{aipproc}

\end{document}